\documentstyle[aps,prb,epsf, multicol]{revtex}
\tighten
\begin{document}
\draft
\newcommand{\bn}{{\bf n}}
\newcommand{\bp}{{\bf p}}
\newcommand{\br}{{\bf r}}
\newcommand{\bq}{{\bf q}}
\newcommand{\bj}{{\bf j}}
\newcommand{\bE}{{\bf E}}
\newcommand{\eps}{\varepsilon}
\newcommand{\la}{\langle}
\newcommand{\ra}{\rangle}
\newcommand{\cK}{{\cal K}}
\newcommand{\cD}{{\cal D}}
\newcommand{\bphi}{\bar\phi_0}
\newcommand{\bpsi}{\bar\psi}
\newcommand{\bF}{\bar F}
\newcommand{\mybeginwide}{
    \end{multicols}\widetext
    \vspace*{-0.2truein}\noindent
    \hrulefill\hspace*{3.6truein}
}
\newcommand{\myendwide}{
    \hspace*{3.6truein}\noindent\hrulefill
    \begin{multicols}{2}\narrowtext\noindent
}

\title{
  Higher  cumulants  of voltage fluctuations in current-biased
  diffusive contacts
}
\author{K.\ E.\ Nagaev
}
\address{
  Institute of Radioengineering and Electronics,
  Russian Academy of Sciences, Mokhovaya ulica 11, 125009 Moscow,
  Russia\\}
\date\today
\maketitle
\bigskip
\begin{abstract}
The third and fourth cumulants of voltage in a current-biased
diffusive metal contact of resistance $R$ are calculated for arbitrary
temperatures and voltages using the semiclassical cascade approach. The
third cumulant equals $e^2R^3I/3$ at high temperatures and $4e^2R^3I/15$
at low temperatures, whereas the fourth cumulant equals $2e^2R^3T/3$ at
high temperatures and $(34/105)e^3R^4I$ at low temperatures.

\end{abstract}
\begin{multicols}{2}
\narrowtext

Recently, higher cumulants of current in mesoscopic conductors
received a significant attention of theorists.\cite{Blanter-00a}
This work was pioneered by Levitov and Lesovik,\cite{Levitov-93}
who found that the charge transmitted through a single-channel
quantum contact at zero temperature is distributed according to a
binomial law. Subsequently, these calculations were extended to
conductors with a large number of quantum channels such as
diffusive wires\cite{Lee-95} and chaotic
cavities.\cite{Jalabert-94,Baranger-94,Brouwer-96} More recently,
the third cumulant of current was calculated for a tunnel contact
with interacting quasiparticles.\cite{Levitov-01} The third and
fourth cumulants of current were also calculated for
diffusive-metal contacts for arbitrary temperatures and
voltages.\cite{Gutman-02,Nagaev-02a}

Common to al these papers was that they considered the
fluctuations of current or charge transmitted through a contact at
a constant voltage drop across it. This allowed the authors to
treat independently the charge transmitted through different
quantum channels and at different energies. The assumption of
constant voltage is justified if the resistance of the external
circuit is much smaller than that of the conductor. In actual
experiments, the opposite relation is quite possible and in this
case one can speak of fluctuations of the voltage drop across the
conductor at a constant current. For a system with an Ohmic
conduction the second cumulant of voltage is just the second
cumulant of current in the voltage-biased regime times $R^2$,
where $R$ is the resistance of the conductor. One might think that
higher cumulants of voltage and current are related in a similar
way, but this is not the case. Very recently Kindermann, Nazarov,
and Beenakker\cite{Kindermann-02} showed that higher cumulants of
voltage and current in the current- and voltage-biased conductors
are not related in such a simple way. In particular, they
calculated the low-temperature third and fourth cumulants of charge
transmitted through a multichannel conductor connected in series
with a macroscopic resistor to a voltage source and found that these
cumulants present nonlinear functions of current cumulants of the
same conductor in the voltage-biased mode.

 The purpose of this paper is to calculate the third
and fourth cumulants of voltage in a current-biased
diffusive-metal contact for arbitrary temperatures and to show how the
recently proposed semiclassical cascade approach
\begin{figure}[t]
\epsfxsize6cm \centerline{\epsffile{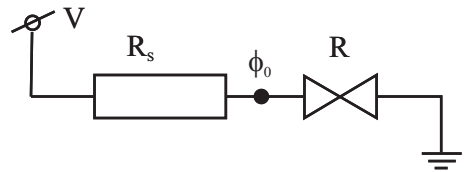}} \vspace{3mm}
\caption{%
 Mesoscopic conductor connected in series with a voltage source
 and an external load.
} \label{fig1}
\end{figure}
\noindent
should be modified in
that case.  The key point for this approach is
that the system is described by at least two distinct variables
whose fluctuations are characterized by essentially different time
scales. Fluctuations of the "slow" variable modulate the intensity
of noise sources for the "fast" variable and result in additional
higher-order correlations. Therefore the higher cumulants of the
fast variable may be recursively expressed in terms of its
lower-order cumulants. Originally, this method was proposed for
diffusive metals.\cite{Nagaev-02a}  Later it was proved to be
equivalent to the rigorous quantum-mechanical approach  for
chaotic cavities.\cite{Nagaev-02b} More recently similar recursive
relations were obtained as a saddle-point expansion of a
stochastic path integral.\cite{Pilgram-02} The
papers\cite{Nagaev-02a,Nagaev-02b} addressed the case of purely elastic
scattering in the low-frequency limit at constant voltage bias where
the fluctuations of the electric potential are inessential and the
cascade expansions can be made with respect to only one parameter, the
distribution function $f(\eps)$. For a current-biased
conductor this is not the case any more and the electric potential
explicitly enters into the expressions. Here we show how the cascade
expansion should be generalized to the case where the system is
described by two different slow variables.

Consider a quasi-one-dimensional diffusive contact of length $L$
and resistance $R$ connected in series with a resistor of larger
cross section that has yet much larger resistance $R_S \gg R$ (see
Fig. 1). Because of a strong energy relaxation in the resistor the
local distribution of electrons is Fermian with a temperature
equal to that of the bath, yet the local electric potential may
fluctuate. Assume that the resistor is connected to the left end
of the contact and the right end of the contact is grounded. A
large constant voltage is applied to the left end of the series
resistor. Therefore the noise of the resistor may be considered as
Gaussian. The fluctuations of current in the circuit are
determined by the larger resistance, hence the spectral density of
the current noise $S_I = 4T/R_S$ is extremely small because of
large $R_S$. Therefore the current through the contact may be
considered as constant even at a nonzero temperature. We will be
interested in fluctuations of the electric potential at the left
end of the contact, which actually presents the voltage drop
across it.

In what follows we consider only fluctuations in the
zero-frequency limit and will neglect the pile-up of charge. It
will be implied that all the subsequent equations contain only
low-frequency Fourier transforms of the corresponding quantities.
A fluctuation of current inside the contact is given by
\begin{equation}
 \delta\bj
 =
 -\sigma\nabla\delta\phi
 +
 \delta\bj^{ext},
 \label{dj}
\end{equation}
where $\sigma$ is the conductivity of the metal, $\delta\phi$ is a
fluctuation of the electric potential, and
\begin{equation}
 \delta\bj^{ext}(\br)
 =
 e N_F
 \int d\eps {\bf \delta F}^{ext}.
 \label{dj^ext}
\end{equation}
The Fourier transform of the correlator of extraneous sources
$\delta{\bf F}^{ext}$ is expressed in terms of the average
distribution function $f$ via a formula\cite{Nagaev-92}
$$
 \la
   \delta F_{\alpha}^{ext}(\eps, \br)
   \delta F_{\beta }^{ext}(\eps', \br')
 \ra
 =
 2\frac{D}{N_F}
 \delta(\br - \br')
$$ \begin{equation}
 \times
 \delta(\eps - \eps')
 \delta_{\alpha\beta}
 f(\eps)[1 - f(\eps)].
 \label{<dF^2>}
\end{equation}
Integrating Eq. (\ref{dj}) over the contact volume, one obtains
\begin{equation}
 \delta I
 =
 \frac{ eN_F}{L}
 \int d^3r \int d\eps\,
 \delta F_x^{ext}
 +
 \frac{1}{R}
 \delta\phi_0.
 \label{dI}
\end{equation}
where $\delta\phi_0$ is a fluctuation of electric potential at the
left end of the contact and $R$ is the contact resistance. As the
current fluctuations are negligibly small, one may set $\delta I =
0$, so that at low frequencies
\begin{equation}
 \delta\phi_0
 =
 -R
 \frac{eN_F}{L}
 \int d^3r \int d\eps\,
 \delta F_x.
 \label{dphi}
\end{equation}
Hence the second cumulant of voltage $\phi_0$ is related to the
second cumulant of current in the voltage-biased contact in a
trivial way
$$
 \la\la
  \phi_0^2
 \ra\ra
 =
 R^2
 \left.
 \la\la
  I^2
 \ra\ra
 \right|_{\phi_0=const},
$$
\begin{equation}
 \la\la
  I^2
 \ra\ra
 =
 \frac{2}{RL}
 \int_0^L dx
 \int d\eps\,
 f(\eps, x)[1 - f(\eps, x)].
 \label{<phi^2>}
\end{equation}

The characteristic time scale for $\delta\bj^{ext}$ and
$\delta{\bf F}^{ext}$ is the elastic scattering time. As this time is
much shorter than the $RC$ time and the time of diffusion across the
contact that describe the evolution of fluctuations of voltage $\delta
\phi_0$ and the distribution function $\delta f(\eps,\br)$. Hence one
may perform  a cascade expansion of higher cumulants with respect to
these slow variables.

 Consider first the third cumulant
of voltage $\la\la\phi_0^3\ra\ra$. Since the third cumulant of
extraneous currents is vanishingly small in the diffusive
limit,\cite{Nagaev-02a} the bare third cumulant of voltage
fluctuations obtained by a direct multiplication of three
equations (\ref{dphi}) is also small. Hence the third cumulant of
voltage should be given by a cascade correction
\begin{equation}
 \la\la
  \phi_0^3
 \ra\ra
 =
 3
 \int_0^L dx \int d\eps
 \frac{
   \delta
   \la\la
     \phi_0^2
   \ra\ra
 }{
   \delta
   f(\eps, x)
 }
 \la
  \delta f(\eps, x)
  \delta\phi_0
 \ra,
 \label{<phi^3>-1}
\end{equation}
where the functional derivative
\begin{equation}
 \frac{
   \delta
   \la\la
     \phi_0^2
   \ra\ra
 }{
   \delta
   f(\eps, x)
 }
 =
 2
 \frac{R}{L}
 [
  1 - 2f(\eps, x)
 ].
 \label{d<phi^2>/df}
\end{equation}
is easily calculated from Eq. (\ref{<phi^2>}). The key difference
from the case of constant voltage is that the fluctuation $\delta
f$ should be calculated now taking into account the feedback from
the environment. The fluctuations of voltage $\phi_0$ caused by
random scattering in the contact result in fluctuations of the
distribution function at the right end of the resistor, which
presents the boundary condition for the distribution function in
the contact. Hence the fluctuation of a distribution function is a sum
\begin{equation}
 \delta f(\eps, x)
 =
 \delta\tilde f(\eps, x)
 +
 \frac{
  \partial f(\eps, x)
 }{
  \partial\phi_0
 }
 \delta\phi_0,
 \label{df}
\end{equation}
where $\delta\phi_0$ is given by Eq. (\ref{dphi}) and
$$
 \delta\tilde f(\eps, x)
 =
 (
  D\nabla^2
 )^{-1}
 \nabla\delta{\bf F}^{ext}
$$
is the ``intrinsic'' part of fluctuation directly caused by random
scattering. The last term
in Eq. (\ref{df}) mixes together fluctuations at different
energies so that they are not independent any more. In the case of
purely elastic scattering in the contact, the average distribution
function $f(\eps, x)$ in the contact is given by
\begin{equation}
 f(\eps, x)
 =
 \psi(x)
 f_0(\eps - e\phi_0)
 +
 \bar\psi(x)
 f_0(\eps)
 \label{f}
\end{equation}
where $\phi_0 = IR$ is the voltage drop across the
contact and $\psi(x) = 1-x/L$ and $\bar\psi(x) = x/L$ are the
characteristic potentials\cite{Buttiker-93} of the left and right
electrodes. Hence the derivative in Eq. (\ref{df}) is just
\begin{equation}
 \frac{
  \partial f(\eps, x)
 }{
  \partial\phi_0
 }
 =
 -e\psi(x)
 \frac{
  \partial f_0
  (
   \eps - e\phi_0
  )
 }{
  \partial\eps
 }.
 \label{df/dphi}
\end{equation}
Multiplying Eqs. (\ref{df}) and (\ref{dphi})  and averaging the
product with use of Eq. (\ref{<dF^2>}), one easily obtains that
\begin{equation}
 \la
  \delta f(\eps, x)
  \delta\phi_0
 \ra
 =
 -2eR
 U(\eps, x)
 +
 R^2
 \frac{
  \partial f(\eps, x)
 }{
  \partial\phi_0
 }
 \la\la
  I^2
 \ra\ra,
 \label{<dfdphi>}
\end{equation}
where
$$
 U(\eps, x)
 =
 \frac{1}{L}
 (
  \nabla^2
 )^{-1}
 \frac{\partial}{\partial x}
 [
  f(1-f)
 ].
$$
A substitution of Eqs. (\ref{d<phi^2>/df}) and (\ref{<dfdphi>})
into Eq. (\ref{<phi^3>-1}) gives
$$
 \la\la
  \phi_0^3
 \ra\ra
 =
 \frac{1}{30}
 eR^2
 \Biggl[
  8T^2
  \sinh
  \left(
   \frac{eIR}{T}
  \right)
  -
  2eIR T
$$ $$
  +
  4eIR T
  \cosh
  \left(
   \frac{eIR}{T}
  \right)
  -
  5e^2I^2R^2
  \coth
  \left(
   \frac{eIR}{2T}
  \right)
 \Biggr]
$$ \begin{equation}
 \Biggl/
 \left[
  T
  \sinh^2
  \left(
   \frac{eIR}{2T}
  \right)
 \right].
 \label{<dphi^3>-3}
\end{equation}
This expression reduces to
$$
 \la\la
  \phi_0^3
 \ra\ra
 =
 \frac{1}{3}
 e^2 R^3 I
$$
at low current or high temperature $eIR \ll T$ and to
$$
 \la\la
  \phi_0^3
 \ra\ra
 =
 \frac{4}{15}
 e^2 R^3 I
$$
at high current or low temperature $eIR \gg T$. In the former case
$\la\la\phi_0^3\ra\ra$ coincides with $-R^3\la\la I^3\ra\ra$, but
in the latter case it is four times larger. On the whole, the
temperature dependence of $\la\la\phi_0^3\ra\ra$ at a given
current appears to be more flat than that of $\la\la I^3\ra\ra$ at
a given voltage. Equation (\ref{<dphi^3>-3}) is in an agreement
with the temperature-dependent third cumulant of voltage obtained
by Beenakker et al.\cite{Beenakker-03} One may also obtain its
low-temperature limit from the formula for the third cumulant of
current of Kindermann et al.\cite{Kindermann-02} using
voltage-biased cumulants for a diffusive contact.\cite{Gutman-02}

Unlike the third cumulant, the fourth cumulant cannot be
expressed in terms of only functional derivatives with respect to
$\delta f$. The point is that the correlator (\ref{<dfdphi>})
explicitly depends on the voltage drop $\phi_0$ through the derivative
$\partial f_0(\eps - e\phi_0)/\partial\eps$. Therefore one has to
perform the cascade expansion with respect to $\delta\tilde f$ and
$\delta\phi_0$ considering them as different stochastic
variables. The rules for constructing the diagrams remain basically
the same as for the case of a voltage-biased contact,\cite{Nagaev-02a}
but now a variation of any fluctuating quantity should be taken twice,
i.e. with respect to $\tilde\delta f(\eps, x)$ and $\delta\phi_0$.
Hence the number of terms significantly increases:
$$
 \la\la
  \phi_0^4
 \ra\ra
 =
 6S_1
 +
 12S_2
 +
 6S_3
 +
 12
 (
  S_4
  +
  S_5
  +
  S_6
  +
  S_7
 )
$$ \begin{equation}
 +
 3S_8
 +
 6S_9
 +
 3S_{10},
 \label{<dphi^4>-1}
\end{equation}
where
$$
 S_1
 =
 \int d\eps_1
 \int d\eps_2
 \int dx_1
 \int dx_2\,
 \frac{
  \delta^2
  \la\la
   \phi_0^2
  \ra\ra
 }{
  \delta f(\eps_1, x_1)
  \delta f(\eps_2, x_2)
 }
$$ \begin{equation}
 \times
 \la
  \delta\tilde f(\eps_1, x_1)
  \delta\phi_0
 \ra
 \la
  \delta\tilde f(\eps_2, x_2)
  \delta\phi_0
 \ra,
 \label{S1-1}
\end{equation}
\begin{equation}
 S_2
 =
 \int d\eps
 \int dx
 \frac{
  \delta^2
  \la\la
   \phi_0^2
  \ra\ra
 }{
  \delta\tilde f(\eps, x)
  \delta\phi_0
 }
 \la
  \delta\tilde f(\eps_1, x_1)
  \delta\phi_0
 \ra
 \la\la
  \phi_0^2
 \ra\ra,
 \label{S2-1}
\end{equation}
\begin{equation}
 S_3
 =
 \frac{
  \partial^2
  \la\la
   \phi_0^2
  \ra\ra
 }{
  \partial
  \phi_0^2
 }
 \la\la
  \phi_0^2
 \ra\ra,
 \label{S3-1}
\end{equation}
$$
 S_4
 =
 \int d\eps_1
 \int d\eps_2
 \int dx_1
 \int dx_2\,
 \frac{
  \delta
  \la\la
   \phi_0^2
  \ra\ra
 }{
  \delta f(\eps_1, x_1)
 }
$$ \begin{equation}
 \times
 \frac{
  \delta
  \la
   \delta\tilde f(\eps_1, x_1)
   \delta\phi_0
  \ra
 }{
  \delta f(\eps_2, x_2)
 }
 \la
  \delta\tilde f(\eps_2, x_2)
  \delta\phi_0
 \ra,
 \label{S4-1}
\end{equation}
\begin{equation}
 S_5
 =
 \int d\eps
 \int dx\,
 \frac{
  \delta
  \la\la
   \phi_0^2
  \ra\ra
 }{
  \delta f(\eps, x)
 }
 \frac{
  \delta
  \la
   \delta\tilde f(\eps, x)
   \delta\phi_0
  \ra
 }{
  \delta\phi_0
 }%
 \la\la
  \phi_0^2
 \ra\ra,
 \label{S5-1}
\end{equation}
\begin{equation}
 S_6
 =
 \frac{
  \partial
  \la\la
   \phi_0^2
  \ra\ra
 }{
  \partial
  \phi_0
 }%
 \int d\eps
 \int dx\,
 \frac{
  \delta
  \la\la
   \phi_0^2
  \ra\ra
 }{
  \delta f(\eps, x)
 }%
 \la
  \delta\tilde f(\eps, x)
  \delta\phi_0
 \ra,
 \label{S6-1}
\end{equation}
\begin{equation}
 S_7
 =
 \left(
  \frac{
   \partial
   \la\la
    \phi_0^2
   \ra\ra
  }{
   \partial
   \phi_0
  }%
 \right)^2
 \la\la
  \phi_0^2
 \ra\ra,
 \label{S7-1}
\end{equation}
$$
 S_8
 =
 \int d\eps_1
 \int d\eps_2
 \int dx_1
 \int dx_2\,
 \frac{
  \delta
  \la\la
   \phi_0^2
  \ra\ra
 }{
  \delta f(\eps_1, x_1)
 }
$$ \begin{equation}
 \times
 \la
  \delta\tilde f(\eps_1, x_1)
  \delta\tilde f(\eps_2, x_2)
 \ra
 \frac{
  \delta
  \la\la
   \phi_0^2
  \ra\ra
 }{
  \delta f(\eps_2, x2)
 },
 \label{S8-1}
\end{equation}
\begin{equation}
 S_9
 =
 \int d\eps
 \int dx\,
 \frac{%
  \delta
  \la\la
    \phi_0^2
  \ra\ra
 }{
  \delta f(\eps, x)
 }%
 \la
  \delta\tilde f(\eps, x)
  \delta\phi_0
 \ra
 \frac{
  \partial
  \la\la
   \phi_0^2
  \ra\ra
 }{
  \partial\phi_0
 },
 \label{S9-1}
\end{equation}
and
\begin{equation}
 S_{10}
 =
 \left(
   \frac{
    \partial
    \la\la
     \phi_0^2
    \ra\ra
   }{
    \partial\phi_0
  }
 \right)^2
 \la\la
  \phi_0^2
 \ra\ra.
 \label{S10-1}
\end{equation}
The numerical prefactors 6, 12, and 3 in Eq. (\ref{<dphi^4>-1})
present the numbers of inequivalent permutations of $\phi_0$ in
the corresponding expressions. The functional derivative with respect
to $\phi_0$ is defined as
$$
 \frac{
  \delta
  \la
   \ldots
  \ra
 }{
  \delta\phi_0
 }%
 =
 \frac{
  \partial
  \la
   \ldots
  \ra
 }{
  \partial\phi_0
 }
 +
 \int d\eps
 \int dx\,
 \frac{
  \delta
  \la
   \ldots
  \ra
 }{
  \delta f(\eps,x)
 }
 \frac{
  \partial f(\eps, x)
 }{
  \partial\phi_0
 }.
$$

The sum $6S_1 + 12S_4 + 3S_8$ gives just $R^4\la\la\tilde I^4 \ra\ra$,
where $\la\la\tilde I^4\ra\ra$ is the fourth cumulant of current for a
current-biased contact with a voltage drop $\la\phi_0\ra$. The sum
$12S_2 + 12S_5$  is easily brought to a form
$$
 12
 \int d\eps
 \int dx\,
 \frac{
  \partial
 }{
  \partial\phi_0
 }%
 \left(
  \frac{
   \delta
   \la\la
    \phi_0^2
   \ra\ra
  }{
   \delta f(\eps, x)
  }%
  \la
   \delta\tilde f(\eps, x)
   \delta\phi_0
  \ra
 \right)
 \la\la
  \phi_0^2
 \ra\ra
$$ $$
 =
 4R^4
 \frac{
  \partial
  \la\la
   \tilde I^3
  \ra\ra
 }{
  \partial\phi_0
 }
 \la\la
  \tilde I^2
 \ra\ra.
$$
The integrals in $S_6$ and $S_9$ also present cumulants $\la\la\tilde
I^3\ra\ra$ for a voltage-biased contact. The whole expression
(\ref{<dphi^4>-1}) assumes the form
$$
 \la\la
  \phi_0^4
 \ra\ra
 =
 R^4
 \la\la
  \tilde I^4
 \ra\ra
 +
 6R^6
 \frac{
  \partial^2
  \la\la
   \tilde I^2
  \ra\ra
 }{
  \partial\phi_0^2
 }%
 \la\la
  \tilde I^2
 \ra\ra^2
$$ $$
 +
 15R^6
 \left(
  \frac{
   \partial
   \la\la
    \tilde I^2
   \ra\ra
  }{
   \partial\phi_0
  }%
 \right)^2
 \la\la
  \tilde I^2
 \ra\ra
 -
 6R^5
 \frac{
  \partial
  \la\la
   \tilde I^2
  \ra\ra
 }{
  \partial\phi_0
 }%
 \la\la
  \tilde I^3
 \ra\ra
$$ \begin{equation}
 -
 4R^5
 \frac{
  \partial
  \la\la
   \tilde I^3
  \ra\ra
 }{
  \partial\phi_0
 }%
 \la\la
  \tilde I^2
 \ra\ra.
 \label{<dphi^4>-2}
\end{equation}
Substituting the cumulants of current for the voltage-biased
contact,\cite{Nagaev-02a} one easily obtains
$$
 \la\la
  \phi_0^4
 \ra\ra
 =
 \frac{1}{2520}
 e^2 R^3
 \Biggl\{
  51
  eIR T^2
  \cosh
  \left(
   \frac{5eIR}{2T}
  \right)
$$ $$
  +
  72 T^3
  \sinh
  \left(
   \frac{5eIR}{2T}
  \right)
$$ $$
  -
  ( 456 T^3 + 224 e^2 I^2R^2 T )
  \sinh
  \left(
   \frac{3eIR}{2T}
  \right)
$$ $$
  +
  ( 70 e^3I^3R^3 - 399 e IR T^2 )
  \cosh
  \left(
   \frac{3eIR}{2T}
  \right)
$$ $$
  +
  1008 T^3
  \sinh
  \left(
   \frac{eIR}{2T}
  \right)
$$ $$
  +
  ( 560 e^3 I^3R^3 + 348 e IR T^2 )
  \cosh
  \left(
   \frac{eIR}{2T}
  \right)
 \Biggr\}
$$ \begin{equation}
 \Biggr/
 \left[
  T^2
  \sinh
  \left(
   \frac{eIR}{2T}
  \right)
 \right]^5.
 \label{<dphi^4>-3}
\end{equation}
This expression reduces to
$$
 \la\la
  \phi_0^4
 \ra\ra
 =
 \frac{2}{3}
 e^2 R^3 T
$$
in the high-temperature limit, which differs from the corresponding
cumulant of current in a voltage-biased contact just by a factor
$R^4$. In the high-current limit,
$$
 \la\la
  \phi_0^4
 \ra\ra
 =
 \frac{34}{105}
 e^3 R^4 I.
$$
This value is in an agreement with the
formula for the low-temperature fourth cumulant of transmitted charge
of Kindermann et al.\cite{Kindermann-02} Unlike the
fourth cumulant of current, the fourth cumulant of voltage is positive
for all currents and
temperatures and its numerical prefactor in the high-current limit is
larger by more than an order of magnitude than that of the fourth
cumulant of current at high voltages.

I am grateful to C. W. J. Beenakker for a discussion. I am also
grateful to S. Pilgram and E. V. Sukhorukov for helping me to find a
term missed in the previous version of the paper.

This work was supported by Russian Foundation for Basic Research,
grant 01-02-17220, and by the INTAS Open grant 2001-1B-14.

\end{multicols}
\end{document}